\documentclass[twocol]{ametsocV6.1}

\usepackage{multirow}

\title{Global kilometer-scale climate simulations: new opportunities for climate services}

\authors{David Saint-Martin\aff{a}\correspondingauthor{David Saint-Martin, david.saint-martin@meteo.fr}, Olivier Geoffroy\aff{a} and Gildas Dayon\aff{a}}
\affiliation{\aff{a}{CNRM, Université de Toulouse, Météo-France, CNRS, Toulouse, France}}

\abstract{Climate services are essential for risk management and economic planning. This paper presents a century-long simulation performed with the ARP-GEM global atmosphere model at 2.6 km horizontal resolution. This simulation was specifically designed to complement and extend the set of simulations used in the latest version of the French climate services. A key feature is its global coverage at kilometer-scale resolution, enabling the representation of phenomena that depend on such fine scales. This is especially relevant for French climate services, as many processes -- particularly over islands and resolution-sensitive regions -- are currently overlooked in global simulations. Additionally, France’s overseas territories are globally dispersed, and this approach allows their representation within a single framework. A limitation of this simulation is the use of prescribed sea surface temperatures due to the lack of ocean coupling; this will be addressed in future work. The present study demonstrates the feasibility of this approach and highlights the benefits of this new generation of climate modeling for climate services.}

\begin{document}

\maketitle

\section{Introduction}

Climate change is a global phenomenon with widespread, observable impacts. Advancing understanding of climate processes and improving projections of future conditions are essential for informing evidence-based adaptation strategies and supporting decision-making across sectors and territories.

The Coupled Model Intercomparison Project Phase 6 (CMIP6) generation of global climate models typically operates at horizontal resolutions of 50 to 100 km \citep[e.g.,][]{eyring-2016}, and their performance thus relies on parameterizations of sub-grid-scale processes, including deep convection, as well as the representation of topography. To refine climate information at regional or local scales, dynamical downscaling techniques can be applied : limited-area models driven by global model outputs produce simulations at finer spatial resolutions on the order of a few kilometers \citep[e.g.,][]{ban-2021}. 

The emergence of a new generation of global climate models operating at kilometer-scale resolution \citep[e.g.,][]{satoh-2008, stevens-2019, freitas-2020, hohenegger-2020, wedi-2020, caldwell-2021, takasuka-2024, geoffroy-2025-1} offers new opportunities to refine climate projections at regional and local scales and improve climate services \citep[e.g.][]{doblas-reyes-2026}. While keeping the added value associated with explicit representation of fine scale processes of the regional models \citep[e.g.,][]{gibson-2024}, global kilometer-scale models provide a major benefit over traditional downscaling approaches : their global coverage allows all regions to be analyzed within a consistent modeling framework, eliminating the need for regionally nested simulations and their associated numerical limitations.

Kilometer-scale resolution enables the explicit representation of key small-scale processes such as deep convection and topography. At grid spacings below $\sim$5 km, a substantial fraction of convective motions can be explicitly resolved, thereby reducing reliance on sub-grid parameterizations and constraining their associated uncertainties. Global kilometer-scale simulations show growing progress in their performance \citep[e.g.,][]{geoffroy-2025-1, takasuka-2026}. They can improve the representation of phenomena such as extreme precipitation \citep[e.g.,][]{stevens-2020}, the diurnal cycle \citep[e.g.,][]{hohenegger-2009, yashiro-2016}, the tropical variability \citep[e.g.,][]{rackow-2025, geoffroy-2025-1}, as well as providing more detailed simulations of cyclones \citep[e.g.,][]{judt-2021}.

Global kilometer-scale simulations pose major challenges in terms of numerical schemes, computational performance, data storage and handling \citep[e.g.,][]{wedi-2014, schulthess-2019}, as well as the representation of remaining subgrid physical processes and the overall model calibration \citep[e.g.,][]{schneider-2024}. Nevertheless, recent advances have demonstrated the technical feasibility of performing global multi-year -- typically 1-10 years -- simulations at kilometer resolution \citep[e.g.,][]{rackow-2025, segura-2025, prein-2026}.

Among these models, the global atmosphere model ARP-GEM \citep{geoffroy-2025} is designed to enable efficient global climate simulations at scales of up to a few kilometers \citep[ARP-GEM2, ][]{geoffroy-2025-1}. In this study, we present an unprecedent century-long global simulation performed with the ARP-GEM2 model at a horizontal resolution of 2.6 km. This simulation was specifically developed to complement the existing simulations integrated into the latest version of the French climate services \citep[DRIAS,][]{lemond-2011}. The global coverage of the ARP-GEM2 simulation is relevant for these services, as it addresses current limitations, especially for islands and regions where climate processes are sensitive to spatial resolution.

Section 2 describes the simulation. Section 3 presents a brief evaluation of its present-day climate and  illustrates projected future climate changes. Conclusions are given in Section 5.

\section{Model and experimental setup}

We use the global atmosphere model ARP-GEM version 2 \citep{geoffroy-2025, geoffroy-2025-1}. The ARP-GEM model is a global, efficient, and multiscale version of ARPEGE/IFS. The model uses a semi-implicit, semi-Lagrangian spectral dynamical core and is formulated under the hydrostatic assumption. This dynamical core, as well as the physical package employed, allows the use of large time steps. In addition, the model benefits from a suite of optimizations, ultimately making the model able of climate simulation at horizontal resolutions ranging from $\mathcal{O}$(100) km up to $\mathcal{O}$(1) km, and allowing calibration at high resolution \citep{geoffroy-2025-1}.

We performed a 124-year global simulation at a horizontal resolution of 2.6 km, covering the period 1976–2099. The ARP-GEM atmospheric model is forced by sea surface temperatures (SSTs) derived from simulations conducted for the CMIP6 project, with the NorESM-MM climate model \citep{seland-2020}. For the period 1976-2014, sea surface temperatures from the CMIP6 {\small \texttt{historical\_r1i1p1f1}} simulation were used, while for 2015–2099, the CMIP6 {\small \texttt{ssp585\_r1i1p1f1}} scenario was applied. We take advantage of the fixed SSTs setup by splitting the 1976-2099 simulation into four segments of about 30 years.

The choice of NorESM-MM was based on a serie of low-resolution (50 km) simulations conducted with the ARP-GEM model, forced by sea surface temperatures derived from multiple CMIP6 models. Among these, the climatology produced using NorESM-MM SSTs showed the best performance, with its global mean surface temperature most closely matching observations. This agreement facilitates the use of existing ARP-GEM model calibrations based on observed SSTs. Furthermore, NorESM-MM has already been used in other experiments conducted by the French climate services, enabling easier comparisons in future studies.

\begin{figure}
\centerline{\includegraphics[width=19pc]{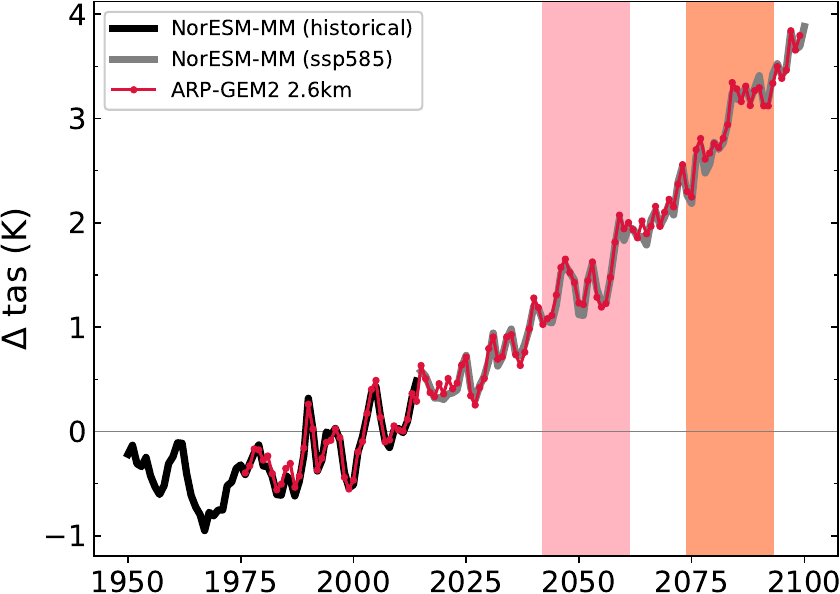}}
\caption{Global surface air temperature changes (unit: K) relative to the 1995–2014 average for the ARP-GEM2 simulation (red line) and the NorESM-MM CMIP6 \textit{historical} (black line) and \textit{ssp585} (gray line) simulations. The 1.5 K GWL corresponds to the 20-year period 2042–2061 (pink shaded band), while the 3 K GWL is reached during the 20-year period 2074–2093 (orange shaded band).}
\label{fig:fig1}
\end{figure}

The temporal evolution of the global surface air temperature change in the NorESM-MM historical and ssp585 simulations is shown in Figure \ref{fig:fig1}. The ARP-GEM2 simulation closely follows the NorESM-MM simulations throughout the period, with small differences arising from the representation of land surfaces. The NorESM model has an equilibrium climate sensitivity of 2.5 K \citep{seland-2020}. Therefore, the ARP-GEM2 simulation enables the sampling of global warming levels (GWLs) ranging from 1 K to 3 K, where a GWL is defined as the period when the 20-year running mean of global surface air temperature exceeds a given threshold relative to the present-day climate. The corresponding periods are 2042-2061 for 1.5 K and 2074-2093 for 3 K.

Given the effectiveness of pattern scaling techniques for estimating climate responses from a reference simulation \citep[e.g.][]{geoffroy-2014}, simulating such a long period is not computationally optimal, as it may introduce redundant information. For example, it would be preferable to simulate only a 20-year period representative of a given GWL and instead sample other sources of uncertainty, such as different SST forcings. Nevertheless, we follow the protocol established by the French climate services to ensure consistency and facilitate comparison with other available datasets. In the future, the use of ocean coupling will be necessary and will require the development of a dedicated protocol to optimize the simulation strategy \citep[e.g.][]{moon-2025} and minimize the number of years to be simulated.

Simulations are performed on ECMWF’s high-performance computing facility, Bull Sequana XH2000\footnote{https://top500.org/system/180214}. Each computational node is equipped with two AMD Epyc Rome processors, each with 64 cores operating at 2.25 GHz. At 2.6 km resolution, an one-year global ARP-GEM simulation can be achieved at a reasonable cost of 650 kh.CPU. This configuration performs 200 simulated days per day (SDPD) over 110 nodes on the ECMWF supercomputer. The 30-year simulations required an elapsed time of approximately two months.

\begin{figure}
\centerline{\includegraphics[width=19pc]{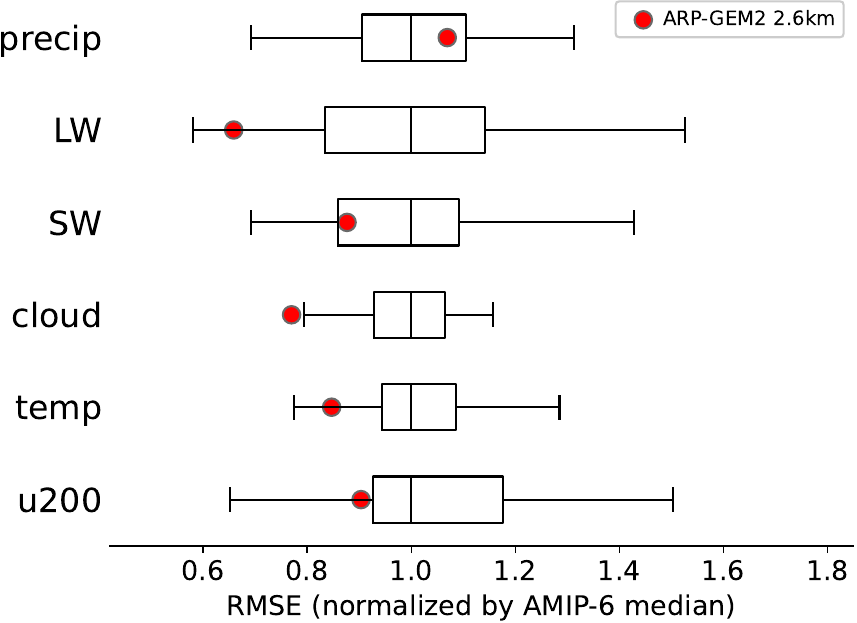}}
\caption{Annual normalized root-mean-square errors (RMSEs) in the climatology of precipitation (Precip), top-of-atmosphere longwave (LW) and net shortwave (SW) radiation, total cloud cover (Cloud), surface air temperature (Temp), and 200-hPa zonal wind (U), calculated against observational or reanalysis datasets. RMSE is normalized by the median value across 40 CMIP6 historical models. These median values for precipitation, LW radiation, SW radiation, total cloud cover, surface air temperature, and 200-hPa zonal wind are for 1985-2014 : 1.1 mm~day$^{-1}$, 8.1 W~m$^{-2}$, 11.5 W~m$^{-2}$, 11.2 \%, 2.5 K, and 2.8 m~s$^{-1}$. RMSEs for ARP-GEM2 at 2.6 km resolution (red dots) and CMIP6 models (box plots) are computed over the 1985–2014 period.}
\label{fig:fig2}
\end{figure}

\section{Present and future climate}

Figure \ref{fig:fig2} shows the root mean square errors (RMSEs) from the ARP-GEM2 simulation, compared to those from the historical CMIP6 simulations for the 1985-2014 climatological means of six key variables: surface temperature, precipitation, net shortwave and longwave radiation at the top of the atmosphere, cloud cover and 200-hPa zonal wind. The ARP-GEM2 simulation at 2.6 km resolution ranks among the best-performing models when compared to the CMIP6 ensemble, indicating that the model is well-calibrated, a particular concern for kilometer-scale models \citep{schneider-2024}. The RMSE for precipitation is slightly higher than that of the model forced by observed sea surface temperatures \citep{geoffroy-2025-1}, suggesting a potential area for future improvement.

\begin{figure*}
\centerline{\includegraphics[width=38pc]{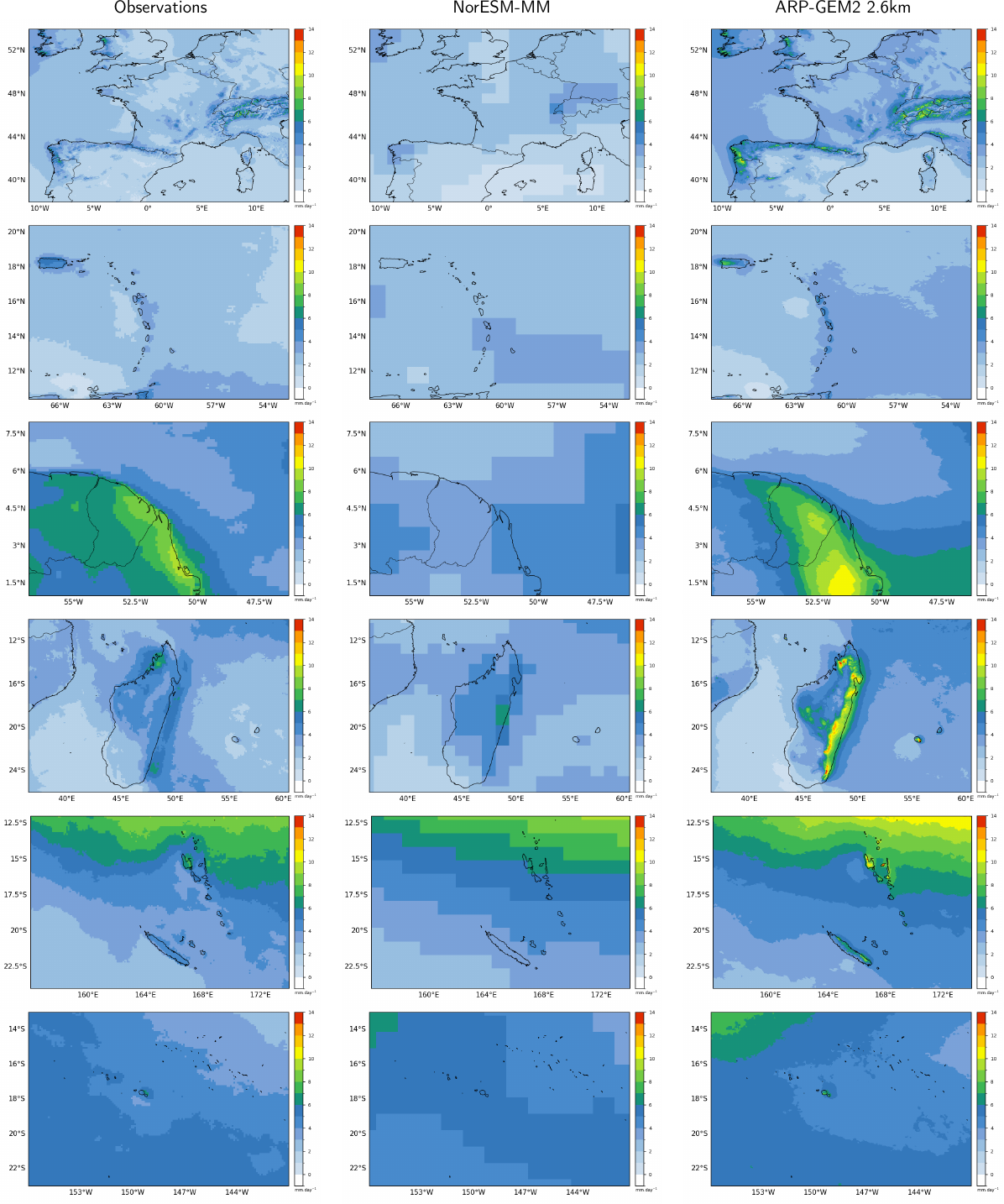}}
\caption{
Mean annual climatology of precipitation (unit: mm.day$^{-1}$) from observational datasets (first column), the NorESM-MM simulation (second column), and the ARP-GEM2 simulation (third column) for six domains covering metropolitan France (first row) and French overseas departments and territories: Guadeloupe and Martinique (second row), French Guiana (third row), Mayotte and Réunion islands (fourth row), New Caledonia (fifth row), and French Polynesia (sixth row). For the Western Europe domain, the climatology is computed over the 1985–2014 period, and observations are derived from the CERRA dataset. For the five other domains, the climatology is computed over the 2001–2020 period, and the observational dataset is IMERG V07B. The areas shown in this figure are indicated by red boxes in Fig. \ref{fig:fig4}.}
\label{fig:fig3}
\end{figure*}

Figure \ref{fig:fig3} compares the annual precipitation climatologies for the present-day climate across the six domain of interest for French climate services: Western Europe, the Caribbean, the Guianas, the Western Indian Ocean region, New Caledonia, and French Polynesia. For the Western Europe domain, climatology is computed for period 1985-2014 and the observations are derived from the Copernicus European Regional Reanalysis datasets \citep[CERRA, ][]{ridal-2024}. For the other five domains, climatology is computed over the period 2001-2020 and the observational dataset used is IMERG V07B \citep{huffman-2019, huffman-2022}.

Across these regions, the ARP-GEM kilometer-scale simulation reproduces the mean precipitation intensity satisfactorily, although it tends to slightly overestimate average rainfall compared to  reference datasets. At the regional scale, the kilometer-scale resolution of ARP-GEM allows it to capture orographic effects (e.g., Cantabrian Mountains, Pyrenees, Alps, Corsica) as well as island and coastal effects. This capability is particularly evident on small islands in the Caribbean, New Caledonia, French Polynesia (e.g., Moorea) or in Indian ocean (e.g. Réunion), but it is also noticeable along coasts such as those of Normandy and the United Kingdom.

\begin{figure*}
\centerline{\includegraphics[width=33pc]{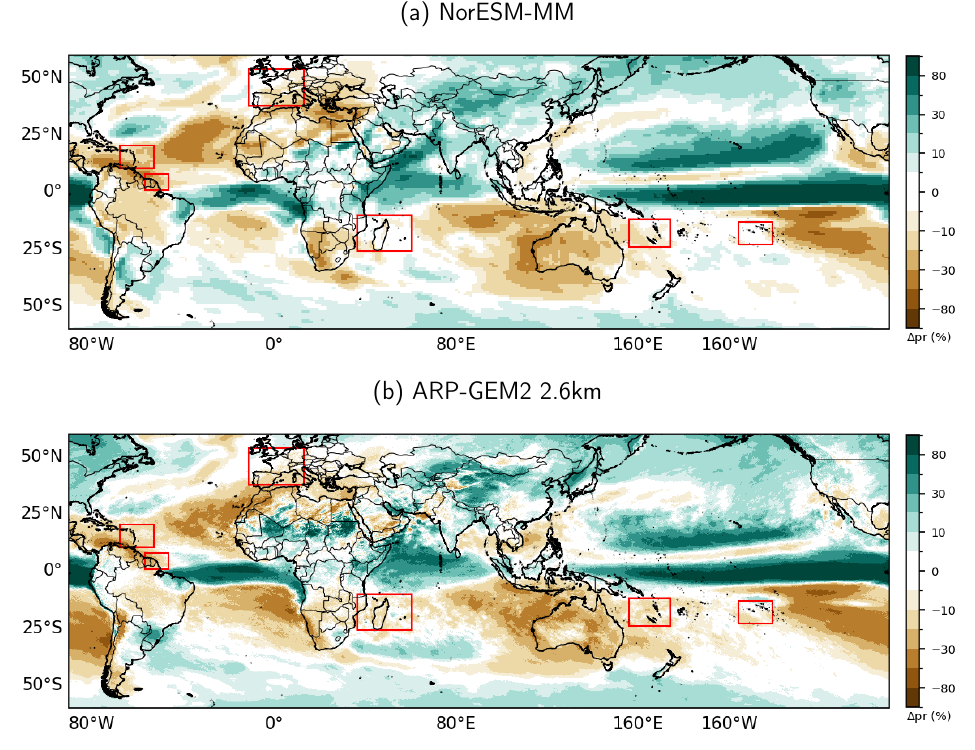}}
\caption{Annual mean precipitation change (\%) in 2064-2099 relative to 1985–2020 for (a) the NorESM-MM model and (b) the ARP-GEM2 at 2.6 km model. Red boxes correspond to the regions plotted in Fig \ref{fig:fig3}.}
\label{fig:fig4}
\end{figure*}

We now turn to the future climate, defined here as the climate simulated by the NorESM-MM and ARP-GEM2 models in response to increases in greenhouse gases and aerosols following the ssp585 warming scenario. Figure \ref{fig:fig4} compares the relative changes in annual precipitation between the future period (2064–2099) and the present period (1985–2020). The overall pattern of changes aligns with expectations from climate models: the wettest regions tend to become wetter, while areas of subsidence and currently dry regions tend to become drier. Both simulations rely on the same prescribed SSTs, which constrain the large-scale climate response. While the broad global response is consistent, differences can be observed between the changes simulated by NorESM-MM and those simulated by ARP-GEM, particularly over French overseas regions. ARP-GEM captures changes over these small islands that are difficult or impossible to anticipate from NorESM-MM alone or even from the CMIP6 multi-model ensemble. 

\section{Conclusion}

This article presents the results of a 124-year simulation at 2.6-kilometer resolution, designed to support French climate services. To our knowledge, no global simulation of such duration at near-kilometer resolution has yet been reported. This simulation was conducted using the global atmosphere model ARP-GEM2, a highly efficient and calibrated model. 

This 2.6 km resolution, century-long global simulation demonstrates the feasibility of such experiments and highlights global kilometer-scale climate modeling as a unique opportunity for delivering climate services.
The simulation required only a small fraction of the computational resources available at an institution like Météo-France -- approximately 2\% of its annual computing budget -- showing that such simulations are no longer restricted to resource-intensive \textit{frontier} experiments, making additional simulations of this type feasible.

The simulation displays particularly interesting features in the mean state and variability of the present climate. Its errors are globally smaller than those of a broad ensemble of state-of-the-art climate models, which typically have horizontal resolutions of $\sim$100 km and are unable to resolve regional-scale structures in key fields such as precipitation. The results support the conclusion that reducing the parameterized component of atmospheric flow, capturing orographic effects, and resolving small islands and coastal regions at high resolution contribute to added value relative to coarser models.

Compared to downscaling approaches, global kilometer-scale simulations offer several advantages. The global system is represented within a consistent framework, ensuring consistency in physics, numerics, and boundary conditions across the entire domain. They explicitly represent deep convection and other fine-scale features, allowing improved scale interactions up to the global scale. Moreover, this unified framework promotes simplicity, ease of use, storage efficiency, and reduced development effort. It can also support collaboration among national climate services through the sharing of climate information from global simulations.

The main limitation of the present simulation is the lack of interactive coupling with an ocean model. The simulation relies on sea surface temperatures from a low-resolution coupled ocean-atmosphere model, which links its climate response to that of the driving model. This aspect will be investigated in future work. Additionally, relying on a single simulation prevents a quantification of model uncertainty. Nevertheless, this climate simulation represents a major step forward for global kilometer-scale climate modeling, offering new opportunities for climate services.

\bibliographystyle{ametsocV6}

\end{document}